\documentclass[amsmath,amssymb,amsfonts,notitlepage,twocolumn]{revtex4}

\usepackage{graphicx}
\usepackage[utf8]{inputenc}
\usepackage[T1]{fontenc}
\usepackage{array}

\begin{document}

\title{Nanofiber Fabry-P\'erot microresonator for non-linear optics and cavity quantum electrodynamics}


\author{C. Wuttke,$^{1}$ M. Becker,$^{2}$ S. Br\"uckner,$^{2}$ M. Rothhardt,$^{2}$ and A. Rauschenbeutel$^{1,*}$}

\address{
$^1$Vienna Center for Quantum Science and Technology, TU Wien -- Atominstitut, Stadionallee 2, 1020 Wien, Austria\\
$^2$Institute of Photonic Technologies  IPHT, Albert-Einstein-Str. 9, 07745 Jena, Germany\\
$^*$Corresponding author: arno.rauschenbeutel@ati.ac.at
}

\begin{abstract}
We experimentally realize a Fabry-P\'erot-type optical microresonator near the cesium D2 line wavelength based on a tapered optical fiber, equipped with two fiber Bragg gratings which enclose a sub-wavelength diameter waist. Owing to the very low taper losses, the finesse of the resonator reaches $\mathcal{F}=86$ while the on-resonance transmission is $T=11$~\%. The characteristics of our resonator fulfill the requirements of non-linear optics and cavity quantum electrodynamics in the strong coupling regime. In combination with its demonstrated ease of use and its advantageous mode geometry, it thus opens a realm of applications. 
\end{abstract}

\maketitle
Efficient coupling of light to quantum emitters, such as atoms, molecules or quantum dots, is one of the great challenges in current research. The interaction can be strongly enhanced by coupling the emitters to the field of subwavelength dielectric waveguides that offer strong lateral confinement of the light. In this context, optical nanofibers, realized as the waist of a tapered optical fiber (TOF), have proven to be a powerful tool \cite{Nayak2008,Morrissey2009,Vetsch2010,Fujiwara2011,Garcia-Fernandez2011}. Another approach towards enhancing light--matter interaction is to employ optical microresonators in which the light is circulating and thus passes the emitters many times \cite{Vahala2003}. 

Here, we present a combination of both approaches and experimentally demonstrate a fully fiber-based optical microresonator which consists of a TOF with two integrated fiber Bragg gratings (FBGs), similar to what has been reported in \cite{Liang2006}. In contrast to other nanofiber FBG cavites\cite{Liang2006,Ding2011b,LeKien2012a}, our microresonator is compatible with the observation of coherent cavity quantum electrodynamics (CQED), such as Rabi oscillations, and its performance is comparable with or superior to other state-of-the-art fiber-based cavities \cite{Colombe2007,Toninelli2010,Goldwin2011}.

\begin{figure}
\centerline{\includegraphics[width=7.4cm]{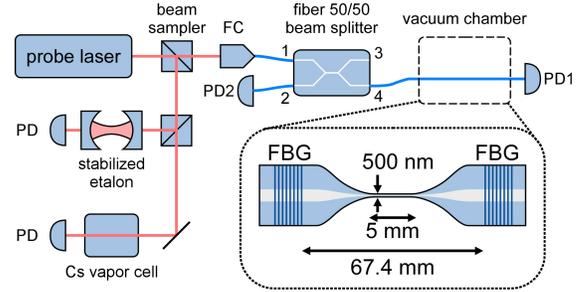}}
\caption{Experimental setup. The inset schematically shows the TOF microresonator. See text for details.}\label{fig1}
\end{figure}

The resonator is schematically depicted in Fig.~\ref{fig1}. It is based on a commercial single mode silica fiber (Fibercore SM800). Two FBGs, separated by $\sim2$~cm, are laser-written into this fiber \cite{Lindner2009} which is then tapered in a heat-and-pull process using a home-built fiber pulling rig. In the resulting resonator, the light is reflected back and forth between the two FBGs. Upon each round-trip, it passes twice the tapered section incorporating the waist with a length of 5~mm and a diameter of 500~nm.

\begin{figure}
\centerline{\includegraphics[width=7.2cm]{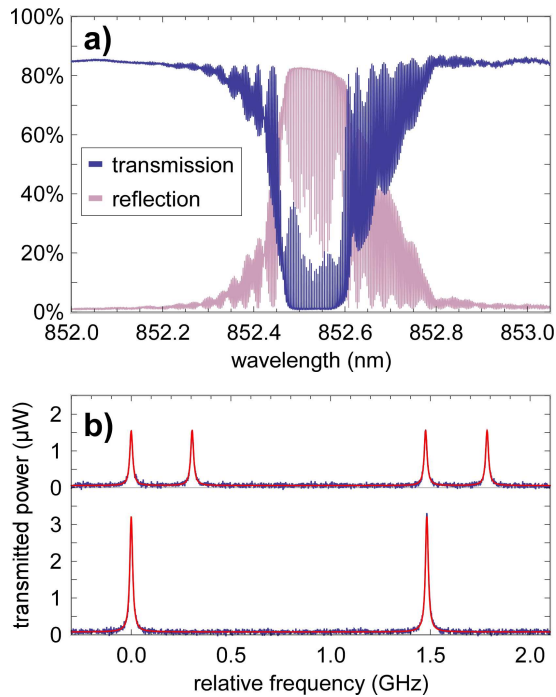}}
\caption{a) Spectral transmission and reflection of the TOF microresonator. Within the 0.2~nm wide stop band, the TOF microresonator modes show up as peaks (dips) in the transmission (reflection). b) Spectral transmission signal over one FSR. The two resonances per FSR in the upper trace correspond to two orthogonal quasi-linearly polarized modes which are split due to intrinsic birefringence. For the lower trace, the polarization of the input light was chosen to match one of these modes. The solid red lines are Airy-function fits.}\label{fig2}
\end{figure}

The experimental setup for the spectral characterization of our TOF microresonator is shown in Fig.~\ref{fig1}. The fiber is placed in a vacuum chamber at a pressure of several $10^{-7}$~mbar in order to test the suitability of the TOF microresonator for cold-atom CQED experiments and to prevent pollution of the fiber waist with dust. A laser beam, derived from an external cavity diode laser with a mode-hop free tuning range of 838--853~nm, is sent through a broadband fiber-based 50/50 beam splitter. Port 4 of the latter is spliced to the fiber that incorporates the TOF microresonator. The measurement protocol consists of the following steps: First, the reflected and transmitted signals are measured with two fiber-coupled photodiodes (PD1 \& PD2). In the following, we remove the TOF microresonator, splice PD1 directly to port 4, and measure the power at port 4 for normalization. Finally, for calibrating the transmission from port 4 to port 2, we splice the fiber coupler (FC) to port 4 and measure the signal on PD2 which is then normalized by splicing the FC to PD2. A cesium vapor absorption spectrum is used as an absolute frequency reference while an etalon is used for relative calibration of the frequency axis. The length of the etalon is actively stabilized using the transmission signal of a second, frequency stabilized laser. The TOF microresonator spectrum is shown in Fig.~\ref{fig2}. 
We fit an Airy-function including an offset to the transmission spectrum and find a maximum finesse of $\mathcal{F}=85.56(32)$ and a free spectral range of $\Delta \nu_{\rm FSR}=1.48084(45)$~GHz at a vacuum wavelength of $\lambda = 852.53(5)$~nm. This is in good agreement with the theoretically expected free spectral range of $\Delta \nu_{\rm FSR}^{\rm th} = c_0/(2 L_{\rm opt}) = 1.47(8)$~GHz, where the optical path length $L_{\rm opt}=10.2(6)$~cm is computed from the known lengths and propagation constants of the different TOF microresonator sections. We also measured the transmission over a broader spectral range than the one shown here using a white light source and a spectrometer with a spectral resolution of 0.2~nm. In the ranges from 700--850~nm and 854--1100~nm, a transmission $T= 90(5)$~\% was observed. The resonator is fixed on a mount with an integrated bending piezo, similar to the one used in \cite{Poellinger2009}. This allows us to tune the resonance frequency of the modes over more than 62~GHz, or 42 FSRs, by mechanically straining the fiber. This technique is also suitable for actively stabilizing the resonance frequency to an external reference \cite{OShea2010}. 

\begin{figure}[b]
\centerline{\includegraphics[width=7.4cm]{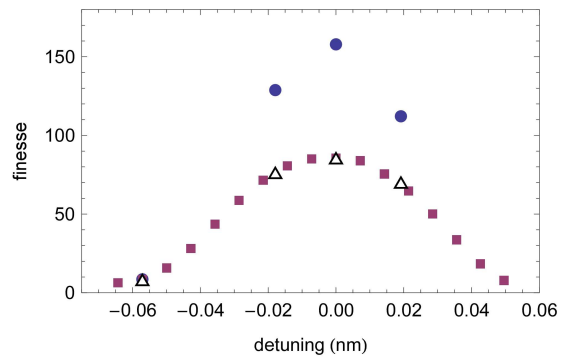}}
\caption{FBG cavity finesses before (circles) and after tapering the fiber (squares) as a function of the laser detuning. Triangles: See text.}\label{fig3}
\end{figure}

Our measurements also allow us to precisely determine the TOF transmission loss. For this purpose, we measure the finesse of the FBG resonator before and after tapering the fiber. In both cases, the finesse is given by
\begin{equation}
	\mathcal{F}=\frac{\pi \sqrt{r_1 r_2 t_c^2}}{1-r_1 r_2 t_c^2}~,
\end{equation}
where $r_1$,$r_2$ and $t_c$ are the amplitude reflection coefficients of the two cavity mirrors and the intracavity amplitude transmission coefficient, respectively. Due to the nearly lossless fiber transmission of the initial fiber, the intracavity losses can be neglected for the  untapered cavity, yielding \mbox{$t_c=1$}. Therefore, the TOF transmission can be calculated from the finesse of the resonator before the tapering, $\mathcal{F}_0$, combined with the finesse of the TOF microresonator, $\mathcal{F}_1$:
\begin{equation}
	t_c^2 = \frac{\mathcal{F}_0^2\left(2 \mathcal{F}_1^2 + \pi^2 -\pi\sqrt{4 \mathcal{F}_1^2 +\pi^2}\right)}{\mathcal{F}_1^2\left(2 \mathcal{F}_0^2 + \pi^2 -\pi\sqrt{4 \mathcal{F}_0^2 +\pi^2}\right)}~.\label{eq:transmTOF}
\end{equation}
Figure~\ref{fig3} shows the cavity finesses before and after the tapering process.
The squares correspond to the finesse after tapering and are determined by fitting Airy functions over one FSR to the data displayed in Fig.~\ref{fig2} a) at the corresponding wavelengths. The detuning is defined as zero at the wavelength of maximum finesse. The circles correspond to the finesse before tapering and are measured by scanning the cavity length via mechanical strain while recording the transmission signal at a fixed laser wavelength. Here, the wavelength is determined using a wavemeter and zero detuning is also defined at the wavelength of maximum finesse. The resulting data was again fitted with an Airy function over one FSR. We chose this method for the untapered FBG resonator because the width of the FBG stop band is on the same order as the FSR of this cavity. A laser frequency scan would not yield reliable results in this situation. 

We infer the TOF transmission by means of Eq.~(\ref{eq:transmTOF}) using the two maximum finesses $\mathcal{F}_0^{\rm max}=158(8)$ and $\mathcal{F}_1^{\rm max}=85.6(6)$ and obtain $t_c^2 = 0.983(1)$. In order to check the validity of this value, we compute the expected finesses after tapering for the remaining data points of the untapered cavity and display them in the same graph (triangles). These projected finesses match the measured values for the TOF microresonator, thereby confirming the consistency of the analysis. We note that the single-pass intracavity losses $1-t_c^2$ are smaller than the total off-resonance TOF-microresonator transmission losses 1-T (see above): The latter includes additional losses due to, e.g., in- and outcoupling through the FBGs.

The capacity of our resonator for enhancing light--matter interaction is characterized by its mode volume in units of the cubic wavelength, $\tilde V$, and its quality factor $Q$ \cite{BermanCQED}. For a given in-coupled power, the resulting intra-cavity intensity is then proportional to $Q/\tilde V.$ In our case, the finesse of $\mathcal{F}_1=86$ in conjunction with the small FSR of $\Delta \nu_{\rm FSR}=1.48$~GHz yields a high quality factor of $Q=2\cdot 10^7$. The normalized mode volume is referenced to the field at the fiber surface and remains as small as $\tilde{V}=2.6\cdot 10^4$, thanks to the strong lateral confinement of the light in the nanofiber section of the TOF. The resulting ratio of $Q/\tilde{V}=783$ thus makes the resonator attractive for non-linear optics applications.

The cavity performance in the context of CQED is governed by the ratios $g/\kappa$ and $g/\gamma$, where $g$ denotes the emitter--light coupling strength, $2\kappa$ the cavity photon decay rate, and $2\gamma$ the free-space spontaneous emission rate of the emitter \cite{BermanCQED}. The conditions for the observation of coherent dynamics then read 
\begin{equation}
g\gg(\kappa,\gamma)~.\label{eq:CQEDcondition}
\end{equation}
These two requirements are often combined in one single (weaker) condition
\begin{equation}
C = g^{2}/2\kappa\gamma\gg1~,\label{eq:CQEDcondition2}
\end{equation}
where $C$ is called cooperativity parameter\cite{BermanCQED}. In the case of a two-level quantum emitter, $C$ is solely determined by cavity properties and is given by 
\begin{equation}
C=\frac{3  \epsilon_0 \mathcal{E}_0^2 \mathcal{F}}{\hbar \beta^3 \Delta \nu_{\rm FSR} }~,
\end{equation}
where $\beta$ is the propagation constant of the mode and $\mathcal{E}_0$ is the field per photon at the position of the emitter. At the fiber surface, we compute a maximum value of $C=29.8(2)$, thereby fulfilling condition (\ref{eq:CQEDcondition2}). 

We now consider the scaling of the CQED parameters with the cavity length $L_{\rm opt}$  \cite{LeKien2009a}. In our case, the intracavity loss, $1-t_c^2$, is governed by non-adiabatic taper losses and is thus largely independent of $L_{\rm opt}$. In this case, $g$ depends on $L_{\rm opt}$ according to $g\propto 1/\sqrt{L_{\rm opt}}$ while $\kappa$ decreases with the cavity length as $\kappa\propto 1/L_{\rm opt}$. Given that $\gamma$ is independent of $L_{\rm opt}$, this yields
$g/\kappa\propto \sqrt{L_{\rm opt}}$, $g/\gamma\propto 1/\sqrt{L_{\rm opt}}$, and $C= {\rm const}$.
Interestingly, the cooperativity parameter is independent of $L_{\rm opt}$. However, the two conditions in Eq.~(\ref{eq:CQEDcondition}) result in two contradicting requirements: $g/\kappa$ is maximized when maximizing $L_{\rm opt}$ while $g/\gamma$ is maximized when minimizing $L_{\rm opt}$. If $C\gg 1$ is fulfilled, we can thus choose the CQED regime that will be realized (e.g., fast cavity regime or coherent dynamics) according to the desired application via the cavity length. We designed our optical path length of $L_{\rm opt}\approx 10$~cm such that we operate in the regime of coherent dynamics: For the $\pi$-polarized Cs D2 transition $\vert F=4,m_{\rm F}=0 \rangle \rightarrow \vert F^\prime=5,m_{\rm F^\prime}=0 \rangle$, we obtain $(g,\kappa,\gamma)/2\pi=(33,~8.6,~2.6)$~MHz. Even with its moderate finesse, the TOF microresonator is thus very well suited for experiments in the coherent CQED regime.

In conclusion, we experimentally realized a Cs D2 line microresonator that fulfills the requirements for non-linear optics applications and for the observation of coherent CQED effects. Our tapered optical fiber microresonator offers advantageous features such as tunabilty, high transmission outside of the fiber Bragg grating stop band and a monolithic design enabling alignment-free operation. Combined with its high coupling strength over the full length of the nanofiber waist, this makes the tapered optical fiber microresonator a promising tool for, e.g., cavity quantum electrodynamics with fiber-coupled atomic ensembles \cite{LeKien2009a, Vetsch2010} and for the realization of quantum network node functionalities, such as, triggered single photon sources \cite{LeKien2011a}, quantum memories based on intra-cavity electromagnetically induced transparency \cite{LeKien2009b}, and entangled two photon sources \cite{LeKien2011b}.

We gratefully acknowledged financial support by the Volkswagen Foundation, and the ESF.

This paper was published in Optics Letters and is made available as an electronic reprint with the permission of OSA. The paper can be found at the following URL on the OSA website: \url{http://dx.doi.org/10.1364/OL.37.001949}. Systematic or multiple reproduction or distribution to multiple locations via electronic or other means is prohibited and is subject to penalties under law.

\end{document}